\documentstyle[sprocl,epsfig,psfig]{article}

\def\Journal#1#2#3#4{{#1} {\bf #2}, #3 (#4)}

\def\PLB{{\em Phys. Lett.}  B}

\def\PRD{{\em Phys. Rev.} D}
\def\EPJ{{\em Eur. Phys. J.} C }
\def\JHEP{{\em JHEP }}

\def\be{\begin{equation}}
\def\ee{\end{equation}}
\def\beq{\begin{equation}}
\def\eeq{\end{equation}}
\def\bea{\begin{eqnarray}}
\def\eea{\end{eqnarray}}
\def\barr{\begin{array}}
\def\earr{\end{array}}
\def\dps{\displaystyle}
\def\ggz{$\gamma\gamma Z$~}
\def\gzz{$\gamma ZZ$~}
\def\eegz{$e^+e^- \to \gamma Z$~}
\def\eett{$e^+e^- \to t \overline t$~}
\def\p{\partial}

\begin{document}

\title{ROLE OF TRANSVERSE POLARIZATION IN CONSTRAINING NEW PHYSICS\\}

\author{SAURABH D. RINDANI }

\address{Theory Group, Physical Research Laboratory\\
Navrangpura, Ahmedabad 380009, India}


\maketitle\abstracts{
Transverse polarization (TP) can be used to study interference of $(S,P)$ or $T$
type couplings from new physics with SM contribution.
In $e^+e^- \to t \bar t$,
a
CP-odd azimuthal asymmetry 
 can constrain 
the scale $\Lambda$ of new $(S,P)$ or $T$  contact interactions
to be higher than about
$
7~{\rm TeV} 
$
for $\sqrt{s}=500$~GeV, $\int {\cal L}dt =  500 {\rm fb}^{-1}$, and assuming 
80\% $e^-$ and 60\% $e^+$ polarizations in opposite directions.
In {\eegz} without chirality violation, but with CP-violating anomalous 
\ggz couplings, an azimuthal asymmetry can probe the anomalous coupling down to about $10^{-2}$.}

Longitudinally polarized beams are expected to be available at a  linear collider.
Spin rotators can then be used to convert longitudinal polarization to transverse
polarization (TP).
Can TP be used to get different information 
as compared to longitudinal polarization?
 Recent work which provides some answers to this
question in various contexts is found in ref. \cite{rizzo}. Here we discuss the question in the  context of some other new physics, mainly from the point of  view of CP violation.

CP violation in $e^+e^- \to f \bar f$ needs 
either polarized $e^+$ and/or $e^-$ beams or measurement of polarization of $f$ and/or $\bar f$.
An exception is 
when $f$ is a neutral particle (like neutralino, $\gamma$,
$Z$, etc), when $\vec{p}_e\cdot\vec{p}_f$ is CP odd, but even under na\" ive 
time reversal.
With TP, one can construct CP-odd triple products which do not need the measurement of the final-state polarization, resulting in better statistics.

Assuming couplings of $e^+e^-$ to new currents of the type $V$, $A$, $S$, $P$ and 
$T$,
it is found that the 
interference of SM $\gamma$ and $Z$ exchange with $V$, $A$
couplings gives 
no CP-violating terms in the distribution, neglecting the electron mass.
$S$ and $T$ couplings do give CP-violating azimuthal terms.
These terms {\em do not need both} $e^-$ and $e^+$ to be polarized.

For $e^-$ and $e^+$ TP's parallel or anti-parallel,
the azimuthal distribution  has 
$\sin 2\phi$ and $\cos  2\phi$ terms, but no
$\sin\phi$ and $\cos\phi$ terms if there is chiral invariance
\cite{hikasa}.
Thus, these latter terms 
can be used to study
interference between SM amplitude and
amplitude with new chirality-violating interactions.
Such terms are absent with longitudinal polarization (or
no polarization).  This is where TP helps.

We illustrate these ideas using the processes 
\eett with contact interactions \cite{prd} and  \eegz with triple-gauge couplings \cite{plb}.
A related discussion on \eett with scalar leptoquark intermediate states may be found in \cite{sdr}.

For the process $e^+e^- \to t \bar t$, a  model-independent four-Fermi Lagrangian for new physics, up to terms of dimension 6, is
\begin{equation}
{\cal L}^{4F}\!
 =\!\!\sum_{i,j=L,R}\Bigl[S_{ij}(\bar{e}P_ie)(\bar{t}P_jt)
 +V_{ij}(\bar{e}\gamma_{\mu}P_ie)(\bar{t}\gamma^{\mu}P_jt)
 +\frac{1}{2}T_{ij}
 (\bar{e}\sigma_{\mu\nu}P_ie)
(\bar{t}\sigma^{\mu\nu}P_jt)\Bigr].
\end{equation}
With TP, the 
 $V$ term does not give CP violation, but the $S$ and $T$ terms do. 
 
The angular distribution with  $S$ and $T$ terms and TP's of  the $e^+$ and $e^-$ of opposite signs has an interference term proportional to $
{\rm Im} S \,\sin\theta\,\sin\phi,
$
which is odd under CP. 
We can then define a CP-odd asymmetry:
\beq\label{asym} A(\theta_0)= \frac{1}{\sigma(\theta_0)} {\dps 
\int_{-\cos\theta_0}^{\cos\theta_0}}d\cos\theta \left[ \dps\int_0^\pi \frac{
d\sigma}{d\Omega} d\phi
 - {\dps\int_{\pi}^{2\pi} \frac{ d\sigma}{d\Omega}} d\phi
\right], \eeq
 where  $\theta_0$ is a cut-off, and $\sigma(\theta_0)$ is the total cross cross section with the cut-off $\theta_0$.
The asymmetry is proportional to the imaginary part of the combination $
S\equiv S_{RR} + \left(2c_A^t c_V^e/ c_V^t c_A^e\right) T_{RR}$.

The result for the 
90\% C.L. limit on Im$S$ 
as a function of the cut-off $\theta_0$
for the choice
 $\sqrt{s}=500$ GeV, 
$\int {\mathcal L} dt = 500$ fb$^{-1}$, and 100\% $e^+$ and $e^-$ polarizations is shown in Fig.~1.  
\begin{figure}[h]
\centering{ \psfig{ file=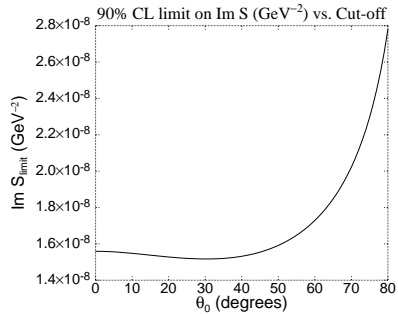,width=1.6in,angle=-90}}
\caption{The 90\%  CL  limit on Im $S$ as a function of the cut-off $\theta_0$.}
\end{figure}
Assuming dimensionless couplings to be ${\mathcal O}(1)$, the limit is
seen to correspond to a new-physics mass scale of $\Lambda \approx 8$ TeV.
With realistic values of polarization, viz., 80\% for $e^-$ and 60\% for $e^+$,
the asymmetry reduces by a factor 0.7,  and the  
limit on $\Lambda$ decreases to about 6.7 TeV. 
With only $e^-$ polarization, the limit decreases to about 5 TeV.

We next consider the process \eegz.
This process has a  neutral final state, and it is possible to study CP violation with TP even without chirality violating interactions.
Asymmetry under  $ \theta \rightarrow \pi - \theta$ 
is a signature of CP violation, which is, however, even under T.
With TP, there is another 
CP-odd, T-odd asymmetry, viz., 
asymmetry under change of sign of the variable
$ \sin^2\theta \cos\theta \sin 2\phi$.
We illustrate this for the concrete case where CP violation arises from anomalous $\gamma\gamma Z$ and $\gamma ZZ$ couplings.

The Lagrangian for anomalous CP-violating couplings of
dimension 6 is
\beq
{\cal L_{\rm anom}} = 
   \frac{e}{m_Z^2}F_{\mu\nu}\left[ \lambda_1 
  \p^\mu Z^\lambda \p_\lambda Z^\nu
               +\frac{1}{16 c_W s_W} \lambda_2
      F^{\nu \lambda}
       \left(\p^\mu Z_\lambda + \p_\lambda Z^\mu   \right)\right].
      \eeq   
The resulting angular distribution of $\gamma$ has terms dependent on Im~$\lambda_1$, Re~$\lambda_2$ and
Im~$\lambda_2$.
We define an asymmetry 
$A_1$ proportinal to Re~$\lambda_2$, corresponding to the difference in the forward-backward asymmetries of events with 
$\sin 2\phi > 0$ and with
$\sin 2\phi < 0$.
The 90\% CL limits that can be obtained on Re$\lambda_2$ using $A_1$ is
Re $\lambda_2<6.2\times 10^{-3}$ assuming a cut-off of 26$^{\circ}$.

To conclude,
TP can be used to study interference of $(S,P)$ or $T$
type couplings from new physics with SM contribution, which is
not possible with longitudinal polarization.
For the case of TP with 
$\vec{s}_{e^-} = - \vec{s}_{e^+}$,
CP-odd asymmetry corresponding to {$\sin\phi \rightarrow  - \sin\phi$}
 can be used to limit new $(S,P)$ or $T$ interactions.
In case of contact $(S,P,T)$ interactions in the process $e^+e^- \to t \bar t$, limits that can be put on the new-physics scale $\Lambda$ is
$\Lambda <
7~{\rm TeV} 
$
for $\sqrt{s}=500$~GeV and $\int {\cal L}dt =  500~{\rm fb}^{-1}$, assuming
80\% $e^-$ polarization and 60\% $e^+$ polarization.

For neutral final states it is possible to have additional CP-violating 
asymmetries with TP even without chirality 
violation. In {\eegz} with CP violation arising from anomalous 
\ggz and \gzz couplings, the
 \ggz coupling gives rise to a CP-violating term proportional to 
{$\sin^2\theta\,\cos\theta\, \sin 2\phi$}
which is CP odd and T odd and hence proportional to 
Re$\lambda_2$. This term can be isolated by a specific asymmetry, and  used to probe Re$\lambda_2$ down to about $10^{-2}$.

\end{document}